\documentclass{svmult}

\usepackage{axodraw}
\usepackage{graphicx}
\usepackage{enumerate}

\def\beq{\begin{equation}}
\def\eeq{\end{equation}}

\def\bea{\begin{eqnarray}}
\def\eea{\end{eqnarray}}
\def\ba{\begin{array}}                  %array
\def\ea{\end{array}}

\begin{document}

\title*{Lepton flavor violation in the SUSY seesaw model: an update}
%\author{Name of Author\inst{1}\and
%Name of Author\inst{2}}
\author{Frank~Deppisch, Heinrich~P\"as, Andreas~Redelbach \and 
Reinhold~R\"uckl}
% Please full name, not just initials!
%\institute{Name and Address of your Institute
%\texttt{name@email.address}
%\and Name and Address of your Institute \texttt{name@email.address}}
\institute{Institut f\"ur Theoretische Physik und Astrophysik\\
Universit\"at W\"urzburg\\ D-97074 W\"urzburg\\ Germany}
\maketitle

We present an update of previous work on charged lepton flavor violation
(LFV) in the seesaw model.
The most recent neutrino data fits and post WMAP mSUGRA benchmark
scenarios are used as input.
In this framework we compare the sensitivity of rare 
radiative decays on the right-handed 
Majorana mass scale \(M_R\) with the reach in slepton-pair production
at a future linear collider.

\section{Introduction}

The evidence for flavor violating neutrino oscillations 
provides strong motivation to
search for LFV also in the charged lepton sector.
While charged lepton-flavor violating processes are suppressed 
in the Standard Model with right-handed neutrinos \cite{petcov}
due to
the light neutrino masses, 
in supersymmetric models new sources of lepton flavor 
violation exist. For example, the massive neutrinos 
affect the renormalization group equations of the slepton masses and 
the trilinear couplings, and give rise to non-diagonal matrix elements 
inducing LFV \cite{Borzumati:1986qx}. This finding
has initiated a prospering research activity both on
LFV in rare decays \cite{rare,Casas:2001sr,Hisano:1999fj} as well as 
on lepton-flavor violating 
processes at future colliders \cite{coll}. 
Recently, we performed a comparative study of
the sensitivity of rare 
radiative decays on the right-handed 
Majorana mass scale \(M_R\) \cite{Deppisch:2002vz}
and the reach in slepton-pair production
at a future linear collider \cite{Deppisch:2003wt}
in the context of the SUSY seesaw model. We  
used mSUGRA benchmark scenarios \cite{Battaglia:2001zp} 
designed for linear collider
studies and paid particular attention 
to the uncertainties in the neutrino input parameters.
Here we present an update of these works, implementing
the most recent neutrino data fits and refined
post WMAP mSUGRA benchmark scenarios.

\section{Neutrino parameters}

\begin{figure}[t!]
\begin{center}
\setlength{\unitlength}{1cm}
%\begin{minipage}[t!]{7.5cm}
\includegraphics[clip,scale=1]{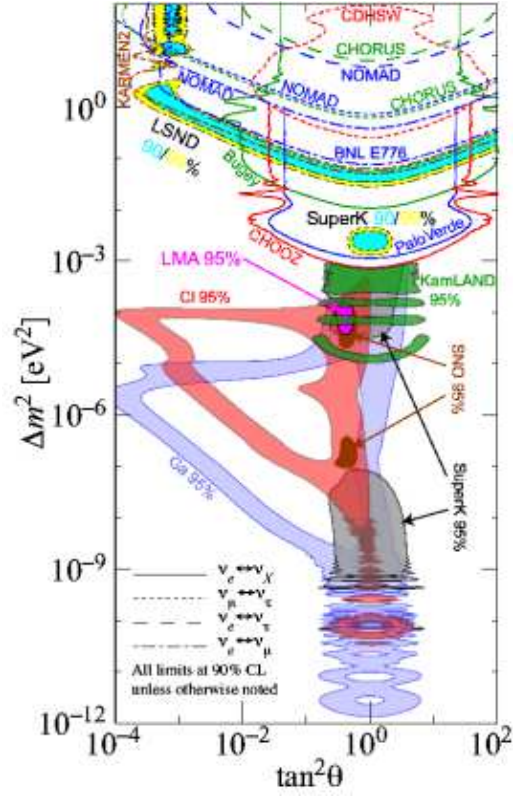}
\end{center}
\caption{Summary of evidences for neutrino oscillations \cite{murayama}.}
     \label{fig:sum}
%\end{minipage}\hfill
%\vspace*{-0.7cm}
\end{figure}

In the last decade a rather unique picture of neutrino 
mixing has emerged. As can be seen in 
Fig.~\ref{fig:sum}, large to maximal mixing has been established for
solar and atmospheric neutrinos, while
the third angle is strongly constrained by reactor measurements.
Recently, this picture of neutrino mixing has been refined further. 
The results of the KamLAND reactor experiment confirmed the disappearance of
solar electron neutrinos, while the SNO experiment allowed for the first time
to study both solar neutrino appearance and disappearance via the separate 
measurement of charged current, elastic scattering and neutral current 
interactions. Moreover, the first data of the K2K long baseline accelerator 
experiment indicate a confirmation of atmospheric neutrino oscillations.
At the time when a linear collider will 
be in operation, even more precise measurements of the
neutrino parameters will be available than today.
In order to simulate the expected improvement, 
we take the central values of the mass squared 
differences $\Delta m^2_{ij}=|m_i^2-m_j^2|$
and mixing angles $\theta_{ij}$ from the most recent global fit to existing 
data \cite{Maltoni:2003da} with errors that indicate 
the anticipated $2\sigma$ intervals of running and proposed 
experiments as further explained in \cite{Deppisch:2002vz}: 
\begin{eqnarray}
&&\tan^2\theta_{23}=1.10^{+1.39}_{-0.60},~~
\tan^2\theta_{13}=0.006^{+0.001}_{-0.006},~~
\tan^2\theta_{12}=0.43^{+0.47}_{-0.22}, \label{nupar1}\\
&&\Delta m_{12}^2=6.9^{+0.36}_{-0.36}\cdot 10^{-5}\textrm{ eV}^2 ,~~
\Delta m_{13}^2=2.6^{+1.2}_{-1.2}\cdot 10^{-3}\textrm{ eV}^2.
\end{eqnarray}
Furthermore, for the lightest neutrino mass we assume 
\(m_1 \leq 0.03\)~eV, where \(m_1 \simeq 0\)~eV corresponds to
the case of a hierarchical spectrum, while \(m_1 \simeq 0.03\)~eV
approaches the degenerate case.

%---------------
%--------------
%\vspace*{2cm}

\section{mSUGRA benchmark scenarios}

For our numerical investigations we
focus on the mSUGRA benchmark scenarios proposed in 
\cite{Battaglia:2003ab} for linear collider studies.
These models are consistent with direct SUSY searches, Higgs searches,
$b \to s \gamma$, and astrophysical constraints.
Recently, the precision measurement of the cosmic microwave background
by the WMAP experiment combined with previously available data provided a 
refined estimate of the density of cold dark matter in the universe. 
Assuming that most of the dark matter is composed of neutralino LSPs, 
the smaller neutralino relic density 
led to a shift of previous benchmark points 
\cite{Battaglia:2001zp}
towards lower values
of the universal scalar mass $m_0$.
In table \ref{mSUGRAscen} we specify the mSUGRA parameters of the
benchmark scenarios B', C', G', and 
I'. These are the only models of \cite{Battaglia:2003ab} with
left-handed sleptons which are light
enough to be pair-produced 
at \(e^+e^-\) colliders with c.m. energies \(\sqrt{s}=500\div800\)~GeV.

\begin{table}[h!]
\begin{center}
\begin{tabular}{|c|c|c|c||c|c|c|}\hline
Scenario & $m_{1/2}$/GeV  & $m_{0}$/GeV & $\tan\beta$ & 
${\tilde m}_{6}$/GeV & 
$m_{\tilde{\chi}_1^0}$/GeV  \\ \hline\hline
B' & 250  & 60  & 10  & 192 & 98  \\ \hline
C' & 400  & 85   & 10  & 291 & 163  \\ \hline
G' & 375  & 115  & 20  & 291 & 153  \\ \hline
I' & 350  & 175  & 35  & 310 & 143  \\ \hline
\end{tabular}
\end{center} 
\caption{\label{mSUGRAscen} Parameters of selected 
mSUGRA post WMAP benchmark scenarios \protect{\cite{Battaglia:2003ab}}. 
The sign of \(\mu\) is chosen to be positive and \(A_0\) is set to 
zero. Given are also the masses of the heaviest
charged slepton and the lightest neutralino, that is the LSP.}
\end{table}

%%%%%%%%%%%%%%%%%%%%%%%%%%%%%%%%%%%%%%%%%%%%%%%%%%%%%%%%%%%%%%%%%%%%%%

\section{SUSY seesaw mechanism and slepton mass matrix}

%In the following we review shortly LFV in the SUSY seesaw model.
If three right-handed neutrino singlet fields $\nu_R$
are added to the MSSM particle 
content, one gets the following additional terms in the superpotential \cite{Casas:2001sr}:
\begin{equation}
W_\nu = -\frac{1}{2}\nu_R^{cT} M \nu_R^c + \nu_R^{cT} Y_\nu L \cdot H_2.
\label{suppot4}
\end{equation}
Here, \(Y_\nu\) is the matrix of neutrino Yukawa couplings, 
$M$ is the right-handed neutrino Majorana mass matrix, and
$L$ and $H_2$ denote the left-handed 
lepton and hypercharge +1/2 Higgs doublets, respectively. 
At energies much below the mass scale of the right-handed neutrinos, 
$W_{\nu}$ leads to the 
mass matrix 
\beq\label{eqn:SeeSawFormula}
M_\nu = m_D^T M^{-1} m_D = Y_\nu^T M^{-1} Y_\nu (v \sin\beta )^2,
\eeq
for the left-handed neutrinos.
Thus, light neutrino masses are 
naturally obtained if the typical scale of the
Majorana mass matrix \(M\) is much larger than the scale of the Dirac mass 
matrix \(m_D=Y_\nu \langle H_2^0 \rangle\),  where
\(\langle H_2^0 \rangle = v\sin\beta\) is the appropriate Higgs v.e.v. with 
\(v=174\)~GeV and 
\(\tan\beta =\frac{\langle H_2^0\rangle}{\langle H_1^0\rangle}\).
Diagonalization of 
$M_{\nu}$ by the unitary MNS matrix 
\begin{eqnarray}
U &=&  \textrm{diag}(e^{i\phi_1}, e^{i\phi_2},1)V(\theta_{12}, 
\theta_{13}, \theta_{23}, \delta), 
\end{eqnarray}
where \(\phi_{1,2}\) and \(\delta\) are the Majorana and Dirac phases, 
respectively, and $\theta_{ij}$ are the mixing angles, 
leads to the 
light neutrino mass eigenvalues $m_i$:
\begin{eqnarray}\label{eqn:NeutrinoDiag}
U^T M_\nu U &=& \textrm{diag}(m_1, m_2, m_3).
\end{eqnarray}

The other neutrino mass eigenstates 
are too heavy to be observed directly. However, they
give rise to virtual corrections 
to the slepton mass matrices 
that can be responsible for observable lepton-flavor violating effects.
In particular, the $6\times 6$ mass matrix of the charged sleptons is given by
\begin{eqnarray}
 m_{\tilde l}^2=\left(
    \begin{array}{cc}
        m_{\tilde l_L}^2    & (m_{\tilde l_{LR}}^{2})^\dagger \\
        m_{\tilde l_{LR}}^2 & m_{\tilde l_R}^2
    \end{array}
      \right)
\end{eqnarray}
with
\begin{eqnarray}
  (m^2_{\tilde{l}_L})_{ab}    &=& (m_{L}^2)_{ab} 
+ \delta_{ab}\bigg(m_{l_a}^2 
+ m_Z^2 
\cos 2\beta \left(-\frac{1}{2}+\sin^2\theta_W \right)\bigg) 
\label{mlcharged} \\
  (m^2_{\tilde{l}_{R}})_{ab}     
&=& (m_{R}^2)_{ab} 
+ \delta_{ab}(m_{l_a}^2 - m_Z^2 \cos 2\beta 
\sin^2\theta_W) \label{mrcharged} \\
 (m^{2}_{\tilde{l}_{LR}})_{ab} 
&=& A_{ab}v\cos\beta-\delta_{ab}m_{l_a}\mu\tan\beta.
\end{eqnarray}
When $m^2_{\tilde{l}}$
is renormalized from the GUT scale \(M_X\) to the electroweak scale 
one obtains, in mSUGRA,
\begin{eqnarray}
m_{L}^2 &=& m_0^2\mathbf{1} + (\delta m_{L}^2)_{\textrm{\tiny MSSM}} + \delta m_{L}^2 \label{left_handed_SSB} \\
m_{R}^2 &=& m_0^2\mathbf{1} + (\delta m_{R}^2)_{\textrm{\tiny MSSM}} + \delta m_{R}^2 \label{right_handed_SSB}\\
A &=& A_0 Y_l+\delta A_{\textrm{\tiny MSSM}}+\delta A \label{A_SSB},
\end{eqnarray}
where $m_{0}$ is the common soft SUSY-breaking scalar mass and $A_{0}$ the 
common trilinear coupling. The terms 
\((\delta m_{L,R}^2)_{\textrm{\tiny MSSM}}\) and 
\(\delta A_{\textrm{\tiny MSSM}}\) are well-known flavor-diagonal corrections.
In addition, the evolution generates off-diagonal terms
in $\delta m_{L,R}^2$ and $\delta A^2$ 
which
in leading-log approximation are given by \cite{Hisano:1999fj}
\begin{eqnarray}\label{eq:rnrges}
  \delta m_{L}^2 &=& -\frac{1}{8 \pi^2}(3m_0^2+A_0^2)(Y_\nu^\dagger L Y_\nu) 
\label{left_handed_SSB2}\\
  \delta m_{R}^2 &=&  0  \\
  \delta A &=&  -\frac{3 A_0}{16\pi^2}(Y_l Y_\nu^\dagger L Y_\nu)
\end{eqnarray}
with
\beq
L_{ij}=\ln\left(\frac{M_X}{M_{i}}\right)\delta_{ij},
\eeq
and $M_i,~i=1,2,3$ being the eigenvalues of the Majorana mass matrix $M$.  
In the above, we have chosen a basis in which 
the charged lepton Yukawa couplings and $M$ are diagonal.

%------------------------------------
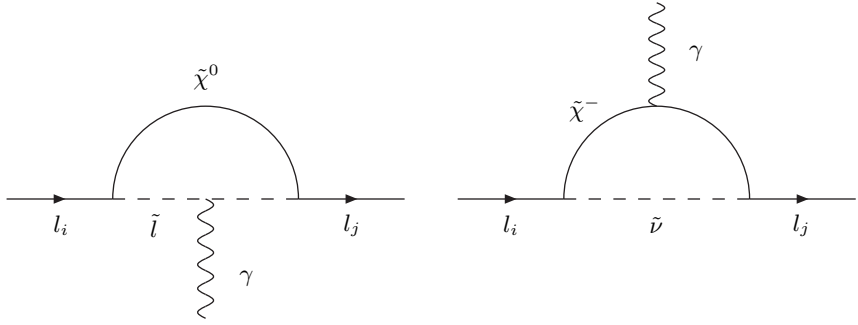
\begin{figure*}[!t]
\begin{center}
\begin{picture}(400,100)(-190,-25)
\ArrowLine(-190,0)(-150,0)
\Text(-170,-10)[]{$l_{i}$}
\DashLine(-150,0)(-80,0){5}
\Text(-135,-10)[]{$\tilde{l}$}
\Photon(-115,0)(-115,-45){3}{5}
\Text(-100,-30)[]{$\gamma$}
\ArrowLine(-80,0)(-40,0)
\Text(-60,-10)[]{$l_{j}$}
\CArc(-115,0)(35,0,180)
\Text(-115,46)[]{$\tilde{\chi}^{0}$}
\ArrowLine(-20,0)(20,0)
\DashLine(20,0)(90,0){5}
\ArrowLine(90,0)(130,0)
\Text(00,-10)[]{$l_{i}$}
\Text(55,-10)[]{$\tilde{\nu}$}
\Photon(55,35)(55,74){3}{5}
\Text(70,55)[]{$\gamma$}
\Text(110,-10)[]{$l_{j}$}
\CArc(55,0)(35,-360,-180)
\Text(28,34)[]{$\tilde{\chi}^{-}$}
\end{picture} 
\vspace*{1cm}
\caption{\label{lfv_lowenergydiagrams} Diagrams for $l_{i}^{-}\rightarrow l_{j}^{-}\gamma$ in the MSSM}
\end{center}
%\vspace*{-0.5cm}
\end{figure*}
The product of the neutrino Yukawa couplings
$Y_\nu^\dagger L Y_\nu$ entering these corrections can be 
determined as follows \cite{Casas:2001sr}. 
By inverting (\ref{eqn:NeutrinoDiag}), one obtains
\begin{eqnarray}\label{eqn:yy}
Y_\nu=\frac{1}{v\sin\beta}\textrm{diag}\left(\sqrt{M_1}, \sqrt{M_2}, \sqrt{M_3}\right) R \; 
\textrm{diag}\left(\sqrt{m_1}, \sqrt{m_2}, \sqrt{m_3}\right)U^\dagger, 
\label{eq:yukawa}
\end{eqnarray}
where $R$ is an unknown  complex orthogonal matrix.
For real $R$ and degenerate right-handed Majorana masses, $R$ as well as 
$\phi_1$ and $\phi_2$ drop out 
from the product $Y_\nu^\dagger L Y_\nu$. 
Using then the neutrino data sketched in section~2
as input the result is evolved
to the unification scale $M_X$.
In what follows we refer to
this illustrative case which suffices for the present discussion.
For small neutrino masses, \(m^2_i \ll \Delta m_{ij}^2\), the above procedure
yields
\begin{eqnarray}
 \left(Y_{\nu}^{\dagger}L Y_{\nu}\right)_{ab}
& \approx & \frac{M_{R}}{v^{2}\sin^{2}\beta}
  \left(\sqrt{\Delta m^{2}_{12}} U_{a2}U_{b2}^{*} 
 + \sqrt{\Delta m_{23}^2}U_{a3}U_{b3}^{*}\right)
 \ln\frac{M_X}{M_{R}}.
\label{llcorrectionhier} 
\end{eqnarray}
Upon diagonalization, the flavor off-diagonal corrections in 
(\ref{left_handed_SSB})-(\ref{A_SSB}) generate flavor-violating couplings
of the slepton mass eigenstates.

\begin{center}
\begin{figure*}[!t]
%--------------------
\vspace*{0.5cm}
\hspace*{-0.5cm}
\( \quad\quad\sum_{a,b} \) 
\hspace*{1cm}
\begin{picture}(220,50)(60,48)
  \ArrowLine(50,50)(10,90)         \Text(10,95)[r]{$e^{+}$}
  \ArrowLine(10,10)(50,50)         \Text(10,5)[r]{$e^{-}$}                          \Vertex(50,50){2}
  \Photon(50,50)(100,50){3}{5}     \Text(75,54)[b]{\(\gamma,Z\)}                    \Vertex(100,50){2}
  \DashArrowLine(130,80)(100,50){5}\Text(113,70)[b]{\(\tilde{l}^{+}_{b}\)}      \Vertex(130,80){2}
  \ArrowLine(160,100)(130,80)      \Text(162,100)[l]{\(l_{j}^{+}\)}
  \ArrowLine(130,80)(160,60)       \Text(162,60)[l]{\(\tilde{\chi}^{0}_{\beta}\)}
  \DashArrowLine(100,50)(130,20){5}\Text(113,30)[t]{\(\tilde{l}_{a}^{-}\)}     \Vertex(130,20){2} 
  \ArrowLine(130,20)(160,40)       \Text(162,40)[l]{\(l^{-}_{i}\)}
  \ArrowLine(160,0)(130,20)        \Text(162,0)[l]{\(\tilde{\chi}_{\alpha}^{0}\)}
\end{picture}
\hspace*{-3.4cm}
\(+~~\sum_{\gamma,a,b}\)
\hspace*{-0.6cm}
\begin{picture}(90,50)(-10,48)
  \ArrowLine(20,80)(0,100)         \Text(0,105)[r]{$e^{+}$}
  \ArrowLine(0,0)(20,20)           \Text(0,-5)[r]{$e^{-}$} 
  \ArrowLine(20,20)(20,80)         \Text(22,50)[l]{\(\tilde\chi_\gamma^{0}\)} \Vertex(20,20){2} \Vertex(20,80){2}
  \DashArrowLine(40,100)(20,80){5} \Text(28,93.5)[b]{\(\tilde{l}_{b}^{+}\)}          \Vertex(40,100){2}
  \ArrowLine(70,100)(40,100)       \Text(72,100)[l]{\(l_{j}^{+}\)}
  \ArrowLine(40,100)(70,70)        \Text(70,70)[l]{\(\tilde{\chi}_{\beta}^{0}\)}
  \DashArrowLine(20,20)(40,0){5}   \Text(28,5)[t]{\(\tilde{l}^{-}_{a}\)}             \Vertex(40,0){2}
  \ArrowLine(40,0)(70,30)          \Text(70,0)[l]{\(\tilde{\chi}_{\alpha}^{0}\)}
  \ArrowLine(70,0)(40,0)           \Text(70,30)[l]{\(l^{-}_{i}\)}
\end{picture}
\vspace{2cm}
\\
\caption{Diagrams for $e^+e^-\to \tilde{l}_b^+\tilde{l}^-_a \to 
l_j^+ l^-_i \tilde{\chi}^0_\alpha\tilde{\chi}^0_\beta$}\label{e+e-_diags}
\end{figure*}
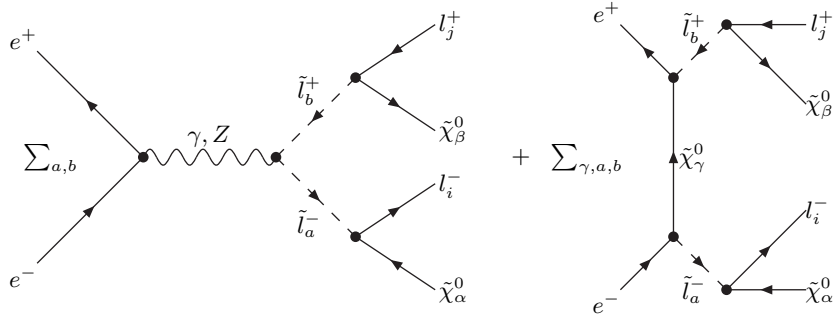
\end{center}
%---------------------------------------------

%%%%%%%%%%%%%%%%%%%%%%%%%%%%%%%%%%%%%%%%%%%%%%%%%%%%%%%%%%%%%%%%%%%%%%%%%%%%%%%
\section{Charged lepton flavor violation}

At low energies, the flavor off-diagonal correction 
(\ref{left_handed_SSB2}) induces
the radiative decays 
\(l_i\rightarrow l_j \gamma\).
From the photon penguin diagrams shown in 
Fig.~\ref{lfv_lowenergydiagrams} with 
charginos/sneutrinos or neutralinos/charged sleptons in the loop, one 
derives the decay rates \cite{Casas:2001sr,Hisano:1999fj}
\begin{equation}
\Gamma(l_i \rightarrow l_j \gamma) 
\propto \alpha^3 m_{l_i}^5 \frac{|(\delta m_L)^2_{ij}|^2}{\tilde{m}^8} 
\tan^2 \beta,
\end{equation}
where $\tilde m$ characterizes the sparticle masses in the loop.
Because of the dominance of the penguin contributions, the process  
\(\mu\to 3e\), and also \(\mu\)-\(e\) 
conversion in nuclei
is directly related to \(\mu\to e \gamma\), e.g.,
\begin{equation}
\frac{Br(\mu\rightarrow 3e)}{Br(\mu\rightarrow e \gamma)} 
\approx \frac{\alpha}{8\pi}\frac{8}{3}
\left(\ln\frac{m_{\mu}^{2}}{m_{e}^{2}}-\frac{11}{4}\right). \label{mu3erel}
\end{equation}

At high energies, a feasible test of LFV is provided 
by the process
$e^+e^- \to \tilde{l}_b^+\tilde{l}^-_a\to 
l_j^+ l^-_i\tilde{\chi}^0_\alpha\tilde{\chi}^0_\beta$.
From the Feynman graphs displayed in Fig.~\ref{e+e-_diags}, one can see that 
LFV can occur in 
production and decay vertices.
For sufficiently narrow slepton 
widths $\Gamma_{\tilde{l}}$ and degenerate masses, the
cross-section can be approximated by
\begin{eqnarray}
\sigma^{\rm pair}_{i \neq j}\propto 
\frac{|(\delta{m}_L)^2_{ij}|^2}{
\tilde{m}^2 \Gamma_{\tilde{l}}^2} \;
\sigma(e^+ e^-\to \tilde{l}^+_j \,
\tilde{l}^-_i) 
Br(\tilde{l}^+_j \to l^+_j \, \tilde{\chi}_0)  
Br(\tilde{l}^-_i \to  l^-_i\, \tilde{\chi}_0)
\label{full_M_squared}.
\end{eqnarray}
where \(\sigma(e^+e^-\to \tilde{l}^+_j \,
\tilde{l}^-_i)\) can be replaced by the 
flavor-diagonal cross-section for slepton pair production. The flavor change
is described by the factor in front of the r.h.s. of (\ref{full_M_squared})
resulting in a single mass insertion.

In the following numerical study we have not assumed
slepton degeneracy 
and have summed the amplitudes 
for the complete \(2 \to 4\) processes
coherently over the intermediate
slepton mass eigenstates. 

%%%%%%%%%%%%%%%%%%%%%%%%%%%%%%%%%%%%%%%%%%%%%%%%%%%%%%%%%%%%%

\section{Numerical Results}

%---------------
\begin{figure}[t!]
\begin{center}
\setlength{\unitlength}{1cm}
%\begin{minipage}[t!]{7.5cm}
\includegraphics[clip,scale=0.65]{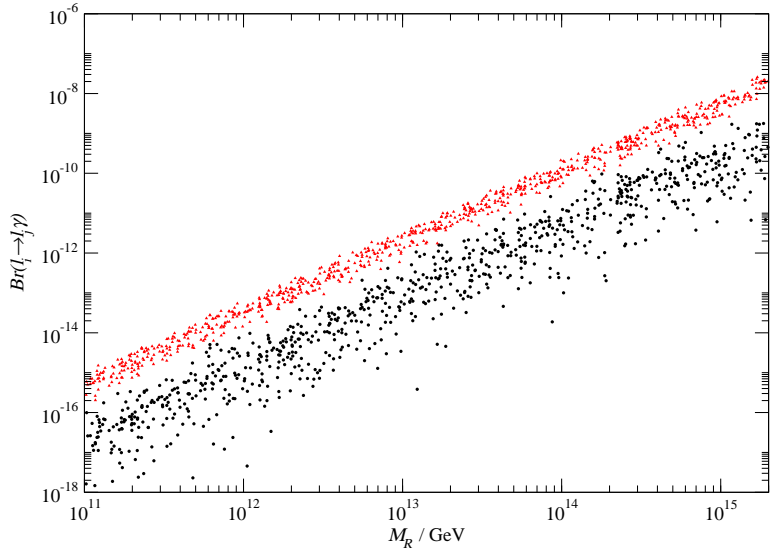}
\end{center}
\caption{Branching ratio of \(\tau \rightarrow \mu\gamma\) (upper band)
and \(\mu \rightarrow e\gamma\) (lower band) in the mSUGRA scenario B'.}
     \label{fig:emu_hier_fut}
%\end{minipage}\hfill
\end{figure}
%------------
\begin{figure}[!t]
\begin{center}
\setlength{\unitlength}{1cm}
%\begin{minipage}{7.5cm}
\includegraphics[clip,scale=0.65]{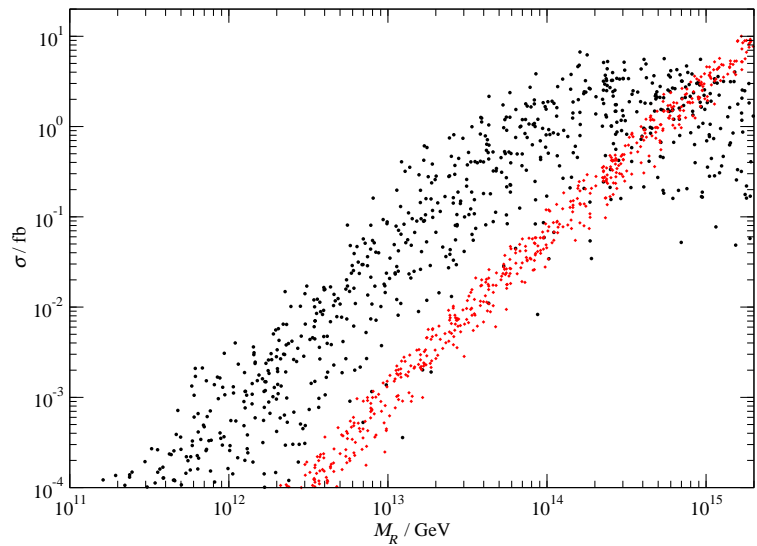}
\end{center}
\caption{Cross-sections for \(e^+e^- \to \mu^+ e^- +2\tilde\chi_1^0\) 
(upper band) and  
\(e^+e^- \to \tau^+ \mu^- +2\tilde\chi_1^0\) (lower band) at
\(\sqrt{s}=500\) GeV
for the mSUGRA scenario B'.
\label{fig:ep}}
%\end{minipage}\hfill
\end{figure}
%-----------

%%%%%%%%%%%%%%%%%%%%%%%%%%%%%%%%%%%%%%%%%%%%%%%%%%%%%%%%%%%%%%%%
Fig.~\ref{fig:emu_hier_fut} illustrates
the dependence of \(Br(\mu\rightarrow e\gamma)\) 
and \(Br(\tau\rightarrow \mu\gamma)\) on the Majorana mass \(M_R\). 
The spread of the predictions reflects the uncertainties in the neutrino 
data. The update of neutrino and SUSY parameters leads to only a slight 
decrease of \(Br(\mu \to e \gamma)\) as compared to the previous results
published in \cite{Deppisch:2002vz}.
One sees that
a branching ratio in the range 
between the present bound and the sensitivity limit of
the new PSI experiment, that is
\(10^{-11} 
\mathrel{\vcenter{\hbox{$>$}\nointerlineskip\hbox{$\sim$}}}
 Br(\mu \to e \gamma) 
\mathrel{\vcenter{\hbox{$>$}\nointerlineskip\hbox{$\sim$}}} 10^{-13}\), 
would point at a value of 
$M_R$ between $5\cdot 10^{12}$~GeV and $5 \cdot 10^{14}$~GeV.
On the other hand,
\(Br(\tau \to \mu \gamma)\) is more strongly affected by
the smaller value of $m_0$ in the new benchmark models 
\cite{Battaglia:2003ab}, resulting in a 
reduction by a factor 
of about 5 as compared to the results in \cite{Deppisch:2002vz}.
Even if \(Br(\tau\to\mu\gamma)=10^{-8}\) is reached, the goal
of SUPERKEKB and LHC \cite{superkekb},
one would only probe $M_R 
\mathrel{\vcenter{\hbox{$>$}\nointerlineskip\hbox{$\sim$}}}
 10^{15}$~GeV
\cite{Deppisch:2002vz}. Nevertheless it is interesting to note
that $\tau \to \mu \gamma$ is much less
affected by the neutrino uncertainties than $\mu \to e \gamma$.

%-----------
\begin{figure}[!t]
\begin{center}
%\begin{minipage}{7.5cm}
\includegraphics[clip,scale=0.65]{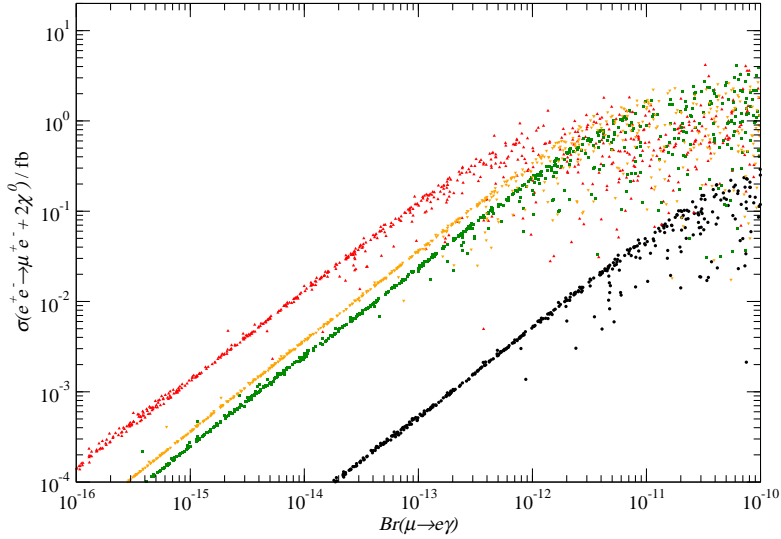}
\end{center}
\caption{Correlation between \(Br(\mu\to e\gamma)\) and   
\(\sigma(e^+e^- \to \mu^+ e^- +
2\tilde\chi_1^0)\) at \(\sqrt{s}=800\) GeV for the mSUGRA 
scenarios (from left to right) C', G', B'  
and I'. \label{fig:emu_lowhigh}}
%\end{minipage}
\end{figure}
%-----------
%-----------
\begin{figure}[!t]
\begin{center}
%\begin{minipage}{7.5cm}
\includegraphics[clip,scale=0.65]{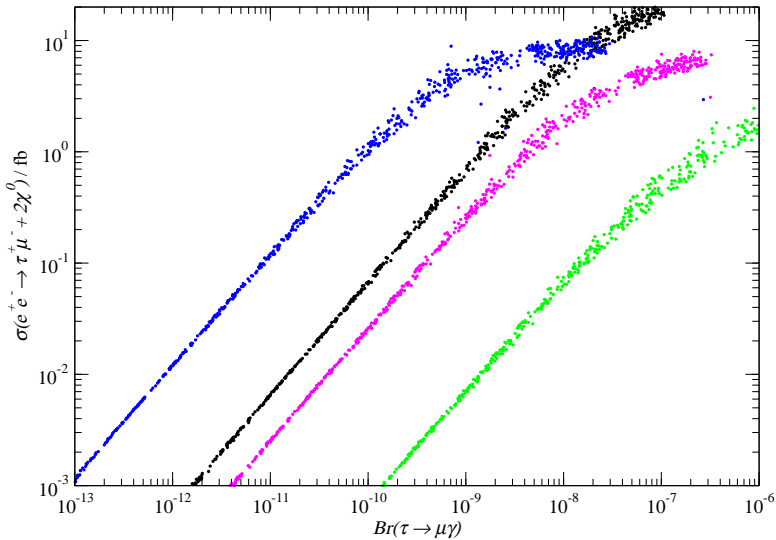}
\end{center}
\caption{Correlation between \(Br(\tau\to \mu\gamma)\) and   
\(\sigma(e^+e^- \to \tau^+ \mu^- +
2\tilde\chi_1^0)\) at \(\sqrt{s}=800\) GeV for the mSUGRA 
scenarios (from left to right)
C', B', G' 
and I'. \label{fig:taumu_lowhigh}}
%\end{minipage}
\end{figure}
%-----------

%%%%%%%%%%%%%%%%%%%%%%%%%%%%%%%%%%%%%%%%%%%%%%%%%%%%%%
Analogously, 
Fig.~\ref{fig:ep} shows the \(M_R\) dependence of the
cross-sections for 
\(e^+e^-\rightarrow \mu^+ e^- +2\tilde{\chi}^0_1\) and 
\(e^+e^-\rightarrow \tau^+ \mu^- +2\tilde{\chi}^0_1\).  
In this case, the
\(\mu e\) channel is enhanced by both the larger $\Delta m_{12}^2$
and the smaller $m_0$ in the new parameter set, resulting in a
cross-section one order of magnitude larger than what was found in
\cite{Deppisch:2003wt}.
For the \(\tau \mu\) final state, on the other hand, the effect of the 
smaller $\Delta m^2_{23}$ is compensated by the enhancement due to 
the smaller value of $m_0$, so that the net effect is negligible.
As can be seen, for a sufficiently large Majorana mass $M_R$ 
the LFV cross-sections can reach several fb. 
The 
\(\tau e\) channel is strongly suppressed by the small mixing angle 
\(\theta_{13}\), and therefore more difficult to observe.

%%%%%%%%%%%Background%%%%%%%%%%%%%%%%%%%%%%%%%%%%%%%%%%%%%%%%%%%%%%%%%%%%%%%%%
The Standard Model background mainly comes from $W$-pair production, 
$W$ production via $t$-channel photon exchange, and $\tau$-pair production.
A 10 degree beam pipe cut and   
cuts on the lepton energy and missing energy reduce the SM background 
cross-sections to less than 30~fb for \(\mu e\) final states and less
than 10~fb for \(\tau \mu\) final states. If one requires a signal
to background ratio $S/\sqrt{S+ B} = 3$,
and assumes an integrated luminosity of 1000~fb$^{-1}$,
a signal cross-section
of 0.1~fb could only afford a background of about 1~fb. 
Whether or not such a low background can be achieved by applying  
selectron selection cuts, for example, on the acoplanarity, 
lepton polar angle and missing transverse momentum has to be studied in a
dedicated simulation. 
For lepton flavor conserving processes one has found
that the SM background 
to slepton pair production can be reduced to about 2-3~fb
at $\sqrt{s}=500$~GeV \cite{Becker:1993fw}.
 
The MSSM background is dominated by
chargino/slepton production with
a total cross-section of
0.2-5~fb and 2-7~fb for \(\mu e \) 
and \(\tau \mu\) final states, 
respectively, depending on the SUSY scenario
and the collider energy. Here, only the direct processes are accounted for.
However, 
the MSSM
background in the $\tau e$ channel can also
contribute to the \(\mu e\)
channel via the decay $\tau \rightarrow \mu \nu_{\mu} {\nu}_{\tau}$.
If $\tilde{\tau}_1$ and $\tilde{\chi}^+_1$ are very light, like in scenarios 
B' 
and I', this background can be as large as 20~fb.
On the other hand, 
such events typically contain two neutrinos in addition to the
two LSPs which are also present in the signal events. Thus, after $\tau$ decay 
one has altogether
six invisible particles instead of two, which may allow to eliminate also
this particularly
dangerous MSSM background 
by cutting on various distributions.
But also here this needs to be studied in a careful simulation.

Particularly
interesting and useful are the 
correlations between LFV in radiative decays and 
slepton pair production. Such a correlation is illustrated in 
Fig.~\ref{fig:emu_lowhigh} for 
\(e^+e^-\rightarrow \mu^+ e^- +2\tilde{\chi}_1^0\)
and \(Br(\mu\to e \gamma)\).
One sees that the neutrino uncertainties 
drop out, 
while the sensitivity to the mSUGRA parameters remains.
Furthermore, while models C', G' and I' are barely affected by the 
change in the new parameter set as compared to the set used in 
\cite{Deppisch:2003wt}, 
in model B' \(\sigma(e^+e^-\rightarrow \mu^+ e^- +2\tilde{\chi}_1^0)\) 
for a given \(Br(\mu \to e \gamma)\)
is by a factor
10 larger than in the previous benchmark point B.
An observation of $\mu \to e \gamma$ with a branching ratio
smaller than $10^{-11}$ would thus be compatible   
with a cross-section as large as 1~fb for 
$e^+e^- \to \sum_{b,a}\tilde{l}_b^+\tilde{l}^-_a\to 
\mu^+ e^- +2\tilde{\chi}^0_1$, at least in model C', G' and B'. 
On the other hand, no signal at the 
future PSI sensitivity of $10^{-13}$
would constrain this channel to less than 0.1~fb.
The correlation of \(Br(\tau \to \mu \gamma)\) and
\(\sigma(e^+e^-\rightarrow \tau^+ \mu^- +2\tilde{\chi}_1^0)\)
is shown in Fig.~\ref{fig:taumu_lowhigh}.
\(Br(\tau \to \mu \gamma)<3 \cdot 10^{-7}\) does not rule out cross-sections
in the $\tau\mu$ channel of 1~fb and larger.
However, one has to keep in mind that these correlations
depend very much on the SUSY scenario.

%%%%%%%%%%%%%%%%%%%%%%%%%%%%%%%%%%%%%%%%%%%%%%%%%%%%%%%%%%%%%%%%%%%%%%%%%%%%%% 
\section{Conclusions}
SUSY seesaw models leading to the observed
neutrino masses and mixings
can be tested by lepton-flavor violating 
processes involving charged leptons.
We have presented an updated analysis of the prospects for 
radiative rare decays 
$l_{i}\rightarrow l_{j}\gamma$ and slepton pair production and decay 
\(e^+e^-\rightarrow \tilde{l}_b^+ \tilde{l}^-_a \rightarrow 
l^+_j l^-_i + /\!\!\!\!E\). 
Assuming the most recent 
global fits to neutrino oscillation experiments \cite{Maltoni:2003da}
we have illustrated the impact of the uncertainties in the 
neutrino parameters. 
Furthermore, using post-WMAP mSUGRA scenarios \cite{Battaglia:2003ab}
we have investigated the
dependence of LFV signals
on the mSUGRA parameters. 
For scenario B' our results can be summarized as follows. 
A measurement
of 
$Br(\mu\rightarrow e\gamma)\approx 
10^{-13}$ 
would probe $M_{R}$ in the range $5 \cdot 10^{12}\div
5 \cdot 10^{13}$~GeV, while
a measurement of $Br(\tau\rightarrow \mu\gamma)
\approx 10^{-8}$ 
would allow to determine  \(M_R \simeq 10^{15}\)~GeV
within a factor of 
2. 
Furthermore, 
\(Br(\mu\rightarrow e \gamma)=10^{-13} \div 10^{-11}\) 
implies  
\(\sigma(e^+e^-\to \mu^+ e^- +2\tilde{\chi}_1^0)=0.02\div 2\)~fb 
at $\sqrt{s}=800$~GeV, while
\(Br(\tau\rightarrow \mu \gamma)=10^{-8} \div 3 \cdot 10^{-7}\) 
predicts
\(\sigma(e^+e^-\to \tau^+ \mu^- +2\tilde{\chi}_1^0)=1\div 10\)~fb, again 
at $\sqrt{s}=800$~GeV.
Hence, linear collider searches are nicely complementary to searches for 
rare decays at low energies and at the LHC.

%Finally, the change in the input parameters relative to 
%\cite{Battaglia:2001zp}, tends to improve the prospects of searches for LFV
%at a future linear collider.

\section*{Acknowledgements}   
This work was supported by the Bundesministerium f\"ur Bildung und 
Forschung (BMBF, Bonn, Germany) under 
the contract number 05HT4WWA2.


\begin{thebibliography}{99}

\bibitem{petcov}
%\cite{Petcov:1976ff}
%\bibitem{Petcov:1976ff}
S.~T.~Petcov,
 %``The Processes Mu $\to$ E Gamma, Mu $\to$ E E Anti-E, Neutrino' $\to$
%Neutrino Gamma In The Weinberg-Salam Model With Neutrino Mixing,''
Sov.\ J.\ Nucl.\ Phys.\  {\bf 25}, 340 (1977)
[Yad.\ Fiz.\  {\bf 25}, 641 (1977\ ERRAT,25,698.1977\ ERRAT,25,1336.1977)].
%%CITATION = SJNCA,25,340;%%
%\cite{Bilenkii:1977du}
S.~M.~Bilenkii, S.~T.~Petcov and B.~Pontecorvo,
%``Lepton mixing, mu --> e + gamma decay and neutrino oscillations,''
Phys.\ Lett.\ B {\bf 67}, 309 (1977).
%%CITATION = PHLTA,B67,309;%%  

%\cite{Borzumati:1986qx}
\bibitem{Borzumati:1986qx}
F.~Borzumati and A.~Masiero,
%``Large Muon And Electron Number Violations In Supergravity Theories,''
Phys.\ Rev.\ Lett.\  {\bf 57}, 961 (1986).
%%CITATION = PRLTA,57,961;%%

\bibitem{rare}
%\cite{Hisano:1996cp}
%\bibitem{Hisano:1996cp}
J.~Hisano, T.~Moroi, K.~Tobe and M.~Yamaguchi,
%``Lepton-Flavor Violation via Right-Handed Neutrino Yukawa Couplings in Supersymmetric Standard Model,''
Phys.\ Rev.\ D {\bf 53}, 2442 (1996)
[arXiv:hep-ph/9510309].
%%CITATION = HEP-PH 9510309;%%
%\bibitem{topdown}
R.~Barbieri, L.~Hall and A.~Strumia,
%``Violations of lepton flavor and CP in supersymmetric unified theories,''
Nucl.\ Phys.\ B {\bf 445}, 219 (1995);
%%CITATION = HEP-PH 9501334;%%
%\cite{Leontaris:1998ue}
G.~K.~Leontaris and N.~D.~Tracas,
%``Lepton flavour violation in unified models with U(1)-family symmetries,''
Phys.\ Lett.\ B {\bf 431}, 90 (1998);
%%CITATION = HEP-PH 9803320;%%
%\cite{Buchmuller:1999gd}
W.~Buchmuller, D.~Delepine and F.~Vissani,
%``Neutrino mixing and the pattern of supersymmetry breaking,''
Phys.\ Lett.\ B {\bf 459}, 171 (1999);
%%CITATION = HEP-PH 9904219;%%
%\cite{Gomez:1999wj}
M.~E.~Gomez, G.~K.~Leontaris, S.~Lola and J.~D.~Vergados,
%``U(1)-textures and lepton flavor violation,''
Phys.\ Rev.\ D {\bf 59}, 116009 (1999);
%%CITATION = HEP-PH 9810291;%%
%\cite{King:1999nv}
S.~F.~King and M.~Oliveira,
%``Lepton flavor violation in string inspired models,''
Phys.\ Rev.\ D {\bf 60}, 035003 (1999);
%%CITATION = HEP-PH 9804283;%%
%\cite{Buchmuller:2000yg}
W.~Buchmuller, D.~Delepine and L.~T.~Handoko,
%``Neutrino mixing and flavor changing processes,''
Nucl.\ Phys.\ B {\bf 576}, 445 (2000);
%%CITATION = HEP-PH 9912317;%%
%\cite{Ellis:2000uq}
J.~Ellis, M.~E.~Gomez, G.~K.~Leontaris, S.~Lola and D.~V.~Nanopoulos,
%``Charged lepton flavour violation in the light of the Super-Kamiokande  data,''
Eur.\ Phys.\ J.\ C {\bf 14}, 319 (2000);
%\cite{Feng:2000wt}
J.~L.~Feng, Y.~Nir and Y.~Shadmi,
%``Neutrino parameters, Abelian flavor symmetries, and charged lepton  flavor violation,''
Phys.\ Rev.\ D {\bf 61}, 113005 (2000);
%%CITATION = HEP-PH 9911370;%%
%\cite{Sato:2000ff}
J.~Sato and K.~Tobe,
%``Neutrino masses and lepton-flavor violation in supersymmetric models  with lopsided Froggatt-Nielsen charges,''
Phys.\ Rev.\ D {\bf 63}, 116010 (2001);
%%CITATION = HEP-PH 0012333;%%
%\cite{Carvalho:2000xg}
D.~F.~Carvalho, M.~E.~Gomez and S.~Khalil,
%``Lepton-flavor violation with non-universal soft terms,''
JHEP {\bf 0107}, 001 (2001).
%\bibitem{bottomup}
%\cite{Sato:2001zh}
J.~Sato, K.~Tobe and T.~Yanagida,
%``A constraint on Yukawa-coupling unification from lepton-flavor  violating processes,''
Phys.\ Lett.\ B {\bf 498}, 189 (2001);
%%CITATION = HEP-PH 0010348;%%;
%\cite{Davidson:2001zk}
S.~Davidson and A.~Ibarra,
%``Determining see-saw parameters from weak scale measurements?,''
JHEP {\bf 0109}, 013 (2001).
%%CITATION = HEP-PH 0104076;%%
%\bibitem{other}
%\cite{Ciafaloni:1996ad}
P.~Ciafaloni, A.~Romanino and A.~Strumia,
%``Lepton flavor violations in SO(10) with large tan beta,''
Nucl.\ Phys.\ B {\bf 458}, 3 (1996);
%%CITATION = HEP-PH 9507379;%%
%\cite{Hisano:1997qq}
J.~Hisano, T.~Moroi, K.~Tobe and M.~Yamaguchi,
%``Exact event rates of lepton flavor violating processes in  supersymmetric SU(5) model,''
Phys.\ Lett.\ B {\bf 391}, 341 (1997);
%%CITATION = HEP-PH 9605296;%%
%\cite{Hisano:1998cx}
J.~Hisano, D.~Nomura, Y.~Okada, Y.~Shimizu and M.~Tanaka,
%``Enhancement of mu --> e gamma in the supersymmetric SU(5) GUT at large  tan(beta),''
Phys.\ Rev.\ D {\bf 58}, 116010 (1998);
%%CITATION = HEP-PH 9805367;%%
%\cite{Hisano:1998tc}
J.~Hisano, D.~Nomura and T.~Yanagida,
%``Atmospheric neutrino oscillation and large lepton flavour violation in  the SUSY SU(5) GUT,''
Phys.\ Lett.\ B {\bf 437}, 351 (1998);
%%CITATION = HEP-PH 9711348;%%
%\cite{Baek:2001kh}
S.~W.~Baek, N.~G.~Deshpande, X.~G.~He and P.~Ko,
%``Muon anomalous g-2 and gauged L(mu) - L(tau) models,''
Phys.\ Rev.\ D {\bf 64}, 055006 (2001);
%%CITATION = HEP-PH 0104141;%%
%\cite{Barenboim:2001ev}
G.~Barenboim, K.~Huitu and M.~Raidal,
%``Flavour violation in SUSY SU(5) GUT at large tan(beta),''
Phys.\ Rev.\ D {\bf 63}, 055006 (2001);
%%CITATION = HEP-PH 0005159;%%
%\cite{Bi:2001tb}
X.~J.~Bi and Y.~B.~Dai;
%``Lepton flavor violation and its constraints on the neutrino mass  models,''
%%CITATION = HEP-PH 0112077;%%
%\cite{Lavignac:2001vp}
S.~Lavignac, I.~Masina and C.~A.~Savoy,
%``tau $\to$ mu gamma and mu $\to$ e gamma as probes of neutrino mass models,''
Phys.\ Lett.\ B {\bf 520}, 269 (2001).
%%CITATION = HEP-PH 0106245;%%
%\cite{Kageyama:2001tn}
%\bibitem{Kageyama:2001tn}
A.~Kageyama, S.~Kaneko, N.~Shimoyama and M.~Tanimoto,
%``Lepton flavor violating processes in bi-maximal texture of neutrino  mixings,''
Phys.\ Rev.\ D {\bf 65} 096010 (2002)
[arXiv:hep-ph/0112359];
%%CITATION = HEP-PH 0112359;%%
%\cite{Babu:2002tb}
%\bibitem{Babu:2002tb}
K.~S.~Babu, B.~Dutta and R.~N.~Mohapatra,
%``Lepton flavor violation and the origin of the seesaw mechanism,''
Phys.\ Rev.\ D {\bf 67}, 076006 (2003)
[arXiv:hep-ph/0211068];
%%CITATION = HEP-PH 0211068;%%
%\cite{Dutta:2003ps}
%\bibitem{Dutta:2003ps}
B.~Dutta and R.~N.~Mohapatra,
%``Lepton flavor violation and neutrino mixings in a 3 x 2 seesaw model,''
Phys.\ Rev.\ D {\bf 68}, 056006 (2003)
[arXiv:hep-ph/0305059].
%%CITATION = HEP-PH 0305059;%%



%\cite{Casas:2001sr}
\bibitem{Casas:2001sr}
J.~A.~Casas and A.~Ibarra,
%``Oscillating neutrinos and mu $\to$ e, gamma,''
Nucl.\ Phys.\ B {\bf 618}, 171 (2001) 
[arXiv:hep-ph/0103065].
%%CITATION = HEP-PH 0103065;%%

%\cite{Hisano:1999fj}
\bibitem{Hisano:1999fj}
J.~Hisano and D.~Nomura,
%``Solar and atmospheric neutrino oscillations and lepton flavor violation  in supersymmetric models with the right-handed neutrinos,''
Phys.\ Rev.\ D {\bf 59}, 116005 (1999)
[arXiv:hep-ph/9810479].
%%CITATION = HEP-PH 9810479;%%


%\cite{Arkani-Hamed:1996au}
\bibitem{coll}
N.~Arkani-Hamed, H.~C.~Cheng, J.~L.~Feng and L.~J.~Hall,
%``Probing Lepton Flavor Violation at Future Colliders,''
Phys.\ Rev.\ Lett.\  {\bf 77}, 1937 (1996)
[arXiv:hep-ph/9603431];
%%CITATION = HEP-PH 9603431;%%
%\cite{Arkani-Hamed:1997km}
%\bibitem{Arkani-Hamed:1997km}
N.~Arkani-Hamed, J.~L.~Feng, L.~J.~Hall and H.~C.~Cheng,
%``CP violation from slepton oscillations at the LHC and NLC,''
Nucl.\ Phys.\ B {\bf 505}, 3 (1997)
[arXiv:hep-ph/9704205];
%%CITATION = HEP-PH 9704205;%%
%\cite{Cheng:1997nd}
%\bibitem{Cheng:1997nd}
H.~C.~Cheng,
%``Flavor and CP violations from sleptons at the muon collider,''
arXiv:hep-ph/9712427;
%%CITATION = HEP-PH 9712427;%%
%\cite{Hirouchi:1997cy}
%\bibitem{Hirouchi:1997cy}
M.~Hirouchi and M.~Tanaka,
%``Lepton flavor violation at linear collider experiments in  supersymmetric grand unified theories,''
Phys.\ Rev.\ D {\bf 58}, 032004 (1998)
[arXiv:hep-ph/9712532];
%%CITATION = HEP-PH 9712532;%%
%\cite{Feng:1998ud}
%\bibitem{Feng:1998ud}
J.~L.~Feng,
%``Supersymmetry at linear colliders: The importance of being e- e-,''
Int.\ J.\ Mod.\ Phys.\ A {\bf 13}, 2319 (1998)
[arXiv:hep-ph/9803319];
%%CITATION = HEP-PH 9803319;%%
%
%\cite{Hisano:1998wn}
%\bibitem{Hisano:1998wn}
J.~Hisano, M.~M.~Nojiri, Y.~Shimizu and M.~Tanaka,
%``Lepton flavor violation in the left-handed slepton production at future  lepton colliders,''
Phys.\ Rev.\ D {\bf 60}, 055008 (1999)
[arXiv:hep-ph/9808410];
%%CITATION = HEP-PH 9808410;%%
%
%\cite{Nomura:2000zb}
%\bibitem{Nomura:2000zb}
D.~Nomura,
%``Probing left-handed slepton flavor mixing at future lepton colliders,''
Phys.\ Rev.\ D {\bf 64}, 075001 (2001)
[arXiv:hep-ph/0004256];
%%CITATION = HEP-PH 0004256;%%
%\cite{Dine:2001cf}
%\bibitem{Dine:2001cf}
M.~Dine, Y.~Grossman and S.~Thomas,
%``Slepton flavor physics at linear colliders,''
eConf {\bf C010630}, P332 (2001)
[Int.\ J.\ Mod.\ Phys.\ A {\bf 18}, 2757 (2003)]
[arXiv:hep-ph/0111154].
%%CITATION = HEP-PH 0111154;%%
%\bibitem{ACFH205}
N.~Arkani-Hamed, J.~L.~Feng, L.~J.~Hall and H.~Cheng,
Nucl.\ Phys.\ B {\bf 505} (1997) 3
[hep-ph/9704205].
%%CITATION = HEP-PH 9704205;%%
%\bibitem{nojiri} 
J.~Hisano, M.~M.~Nojiri, Y.~Shimizu and M.~Tanaka,
Phys.\ Rev.\ D {\bf 60} (1999) 055008
[hep-ph/9808410].
%%CITATION = HEP-PH 9808410;%%
%\bibitem{GKR}
M.~Guchait, J.~Kalinowski and P.~Roy,
%``Supersymmetric lepton flavor violation in a linear collider: The
%role  of charginos,'' 
Eur.\ Phys.\ J.\ C {\bf 21} (2001) 163
[arXiv:hep-ph/0103161].
%%CITATION = HEP-PH 0103161;%%
%\bibitem{PM} W.~Porod and W.~Majerotto,
%``Large lepton flavor violating signals in supersymmetric particle
%decays  at future e+ e- colliders,'' 
Phys.\ Rev.\ D {\bf 66} (2002) 015003
[arXiv:hep-ph/0201284].
%%CITATION = HEP-PH 0201284;%%
%\bibitem{jk01} J.~Kalinowski, 
%``Supersymmetric Lepton Flavor Violation At E+ E- Linear Colliders,''
Acta Phys.\ Polon.\ B {\bf 32} (2001) 3755.
%%CITATION = APPOA,B32,3755;%%
%\bibitem{jk02} J.~Kalinowski, 
%``Slepton flavour violation at colliders,''
Acta Phys.\ Polon.\ B {\bf 33} (2002) 2613
[arXiv:hep-ph/0207051].
%%CITATION = HEP-PH 0207051;%%
%%\cite{Carvalho:2001ex}
%\bibitem{Carvalho:2001ex}
%D.~F.~Carvalho, J.~R.~Ellis, M.~E.~Gomez and S.~Lola,
%%``Charged-lepton-flavour violation in the CMSSM in view of the muon  anomalous magnetic moment,''
%Phys.\ Lett.\ B {\bf 515}, 323 (2001)
%[arXiv:hep-ph/0103256].
%%%CITATION = HEP-PH 0103256;%%
%\cite{Cao:1998kh}
%\bibitem{Cao:1998kh}
J.~J.~Cao, T.~Han, X.~Zhang and G.~R.~Lu,
%``Slepton oscillation at e gamma colliders,''
Phys.\ Rev.\ D {\bf 59}, 095001 (1999)
[arXiv:hep-ph/9808466].
%%CITATION = HEP-PH 9808466;%%
%\cite{Cao:2003zv}
%\bibitem{Cao:2003zv}
J.~Cao, Z.~Xiong and J.~M.~Yang,
%``Lepton flavor violating Z decays in supersymmetric see-saw model,''
Eur.\ Phys.\ J.\ C {\bf 32}, 245 (2004)
[arXiv:hep-ph/0307126].
%%CITATION = HEP-PH 0307126;%%
%\cite{Cannoni:2003my}
%\bibitem{Cannoni:2003my}
M.~Cannoni, S.~Kolb and O.~Panella,
 %``Lepton flavour violation in e+- e- $\to$ l+- e- (l = mu, tau) induced by
%R-conserving supersymmetry,''
Phys.\ Rev.\ D {\bf 68}, 096002 (2003)
[arXiv:hep-ph/0306170].
%%CITATION = HEP-PH 0306170;%%

%\cite{Deppisch:2002vz}
\bibitem{Deppisch:2002vz}
F.~Deppisch, H.~P\"as, A.~Redelbach, R.~Ruckl and Y.~Shimizu,
%``Probing the Majorana mass scale of right-handed neutrinos in mSUGRA,''
Eur.\ Phys.\ J.\ C {\bf 28}, 365 (2003)
[arXiv:hep-ph/0206122].
%%CITATION = HEP-PH 0206122;%%

\bibitem{Deppisch:2003wt}
F.~Deppisch, H.~P\"as, A.~Redelbach, R.~R\"uckl and Y.~Shimizu,
%``The SUSY seesaw model and lepton-flavor violation at a future
%electron positron linear collider,'' 
accepted for publication in Phys. Rev. D,
arXiv:hep-ph/0310053.
%%CITATION = HEP-PH 0310053;%%


\bibitem{murayama}
Homepage Hitoshi Murayama, 
http://hitoshi.berkeley.edu/neutrino/.

%\cite{Maltoni:2003da}
\bibitem{Maltoni:2003da}
M.~Maltoni, T.~Schwetz, M.~A.~Tortola and J.~W.~F.~Valle,
%``Status of three-neutrino oscillations after the SNO-salt data,''
Phys.\ Rev.\ D {\bf 68}, 113010 (2003)
[arXiv:hep-ph/0309130].
%%CITATION = HEP-PH 0309130;%%


%\cite{Battaglia:2003ab}
\bibitem{Battaglia:2003ab}
M.~Battaglia, A.~De Roeck, J.~R.~Ellis, F.~Gianotti, K.~A.~Olive and L.~Pape,
%``Updated post-WMAP benchmarks for supersymmetry,''
arXiv:hep-ph/0306219.
%%CITATION = HEP-PH 0306219;%%

%\cite{Battaglia:2001zp}
\bibitem{Battaglia:2001zp}
M.~Battaglia {\it et al.},
%``Proposed post-LEP benchmarks for supersymmetry,''
Eur.\ Phys.\ J.\ C {\bf 22} (2001) 535
[arXiv:hep-ph/0106204].
%%CITATION = HEP-PH 0106204;%%


\bibitem{superkekb}
L. Serin and R. Stroynowski, ATLAS Internal Note (1997);
T. Ohshima, talk at the 3rd Workshop on Neutrino Oscillations and their 
Origin (NOON2001), 2001, ICRR, Univ. of Tokyo, Kashiwa, Japan;
D. Denegri, private communication.

%\cite{Becker:1993fw}
\bibitem{Becker:1993fw}
R.~Becker and C.~Vander Velde,
%``Aspects of scalar lepton search,''
IIHE-93-08
%\href{http://www.slac.stanford.edu/spires/find/hep/www?r=iihe-93-08}{SPIRES
%entry} 
{\it Prepared for Working Group on e+ e- Collisions at 500-GeV: The
  Physics Potential, Munich, Germany, 20 Nov 1992}.

\end{thebibliography}
\end{document}